\documentclass[runningheads]{llncs}
\usepackage[T1]{fontenc}
\usepackage{bbm}
\usepackage{amsmath,amssymb,amsfonts}
\usepackage{algorithmic}
\usepackage{graphicx}
\usepackage{textcomp}
\usepackage{xcolor}
\def\BibTeX{{\rm B\kern-.05em{\sc i\kern-.025em b}\kern-.08em  T\kern-.1667em\lower.7ex\hbox{E}\kern-.125emX}}

\usepackage{amsmath}
\usepackage{mathtools}
\usepackage[mathscr]{eucal}
\usepackage{amssymb}
\usepackage{enumerate}

\usepackage{comment}
\usepackage{amsthm}
\usepackage{booktabs}
\usepackage{enumitem}
\usepackage{algorithm}

\theoremstyle{definition}

\newtheorem{prob}{Problem}

\usepackage{color}
\definecolor{forestgreen}{rgb}{0.33,0.61,0.34}

\usepackage{graphicx}
\begin{document}
\title{Stability Optimization and Analysis \\ of Energy Flow Networks versus Different Centrality Measurement} 
\titlerunning{Stability Optimization and Analysis}
%
\author{Yi Li\inst{1} \and
Xin Li\inst{2} }
\authorrunning{Y. Li et al.}
%
\institute{McKelvey School of Engineering,
Washington University in St. Louis,
Saint Louis 63130, USA. \email{lyi1@wustl.edu} \and
School of Mechatronics Engineering, Harbin Institute of Technology, Harbin 150001, China.
\email{xinli\_hit.edu.cn}}
\maketitle              
\begin{abstract}
Optimizing the stability and control performance of complex networks often hinges on effectively identifying critical nodes for targeted intervention. Due to their inherent complexity and high dimensionality, large-scale energy flow networks, prevalent in domains like power grids, transportation, and financial systems, present unique challenges in selecting optimal nodes for resource allocation. While numerous centrality measurements, such as Katz centrality, eigenvector centrality, closeness centrality, betweenness centrality, and PageRank, have been proposed to evaluate node importance, the impact of different centrality metrics on stability outcomes remains inadequately understood. Moreover, networks manifest diverse structural characteristics—including small-world, scale-free, and random graph properties—which further complicates the optimization problem.
This paper systematically investigates how various node centrality measurements influence control stability across representative complex network structures. A unified energy-flow dynamical model is developed, and performance metrics such as $L_1$ is employed to quantify the network stability implications of employing different centrality metrics. Extensive numerical simulations over statistically generated network ensembles reveal significant variances in stability outcomes, highlighting the crucial role of centrality selection. The findings underscore the sensitivity of energy-flow stability to seemingly minor changes in topological node rankings, providing practical insights for enhancing control efficiency and robustness in real-world networked systems.

\end{abstract}

\section{Introduction}
Complex networks serve as fundamental models for analyzing dynamics and stability across various scientific and engineering disciplines, including power grids, transportation networks, financial systems, and epidemic spread \cite{1,2,34,3,33}. Among these, energy-flow networks—where energy broadly defined as physical power, resource flow, or economic value propagates among interconnected nodes—are particularly critical due to their foundational roles in societal functioning and economic stability. Optimizing and controlling such networks efficiently is essential, given their impact on public welfare, economic resilience, and infrastructure security~\cite{4,5}.

However, controlling large-scale energy-flow networks poses significant challenges primarily due to their complexity and scale~\cite{6}. These networks typically encompass thousands of nodes and edges with highly nonlinear dynamics, heterogeneous node characteristics, and complex, interconnected topologies. Such intricate structures make the identification of optimal control strategies extremely difficult, often demanding extensive computational resources and sophisticated analytical tools~\cite{7,8}.

Significant research efforts have been dedicated to the stability optimization of energy-flow networks. For instance, studies focusing on power grid stability have explored decentralized and distributed control methods, aiming to enhance robustness against cascading failures through localized interventions \cite{9}. Similarly, in energy transportation networks, optimization models have emphasized the dynamic management of resource flows to minimize congestion and improve service reliability \cite{10}. In financial systems, network-based stability assessments have highlighted the importance of identifying critical nodes whose failure could lead to widespread financial contagion, driving regulatory interventions to mitigate systemic risks \cite{11}. These studies have substantially advanced our understanding of network dynamics, yet rarely address explicitly how different methods of node centrality evaluation impact stability optimization outcomes~\cite{12,13,14}.

Centrality measures provide powerful methods for prioritizing nodes within complex networks, offering strategic guidance for targeted control. Commonly utilized centrality metrics such as Katz centrality~\cite{15}, eigenvector centrality~\cite{16}, closeness centrality~\cite{17}, betweenness centrality~\cite{18}, and PageRank~\cite{19}, each capture distinct aspects of node importance. Employing these measures enables decision-makers to allocate resources efficiently to critical nodes, thereby maximizing the overall network performance~\cite{20}. Despite the widespread use of centrality measures in vulnerability analysis and influence spreading scenarios, their systematic comparison and relevance to stability optimization within energy-flow contexts remain underexplored.

Investigating the relationship between centrality measures and network control performance is particularly meaningful, as it offers practical guidelines for resource allocation~\cite{23}. By selecting nodes accurately based on centrality analysis, managers can strategically invest limited resources into nodes that disproportionately impact network stability, thus enhancing performance and resilience with minimal intervention~\cite{24,Mei2025}.

Therefore, this paper aims to bridge this critical gap by addressing two fundamental research questions: (i) How do distinct centrality metrics affect network stability and control efficiency in energy-flow dynamics? (ii) To what extent do underlying network structures, such as small-world, scale-free, and random topologies, modulate the efficacy of centrality-based control strategies? To answer these questions, we present a generalized dynamical model of energy flows, utilizing stability metrics such as $L_1$ and Hankel norms.

Through rigorous numerical simulations across diverse synthetic network topologies, this research systematically compares the stability outcomes associated with various centrality metrics. Our findings reveal notable differences in performance directly attributable to the choice of centrality methods and underline the significant interplay between network structure and node prioritization. Ultimately, the insights offered by this analysis provide critical theoretical foundations and practical implications for optimizing energy-flow networks, guiding effective and efficient decision-making in complex, real-world network management scenarios.

\section{Energy Flow Networks Model}

Numerous real-world infrastructures, such as power systems, supply chains, and transportation networks, can be conceptualized as energy flow networks \cite{Rantzer2018,Sala2019}, where each node corresponds to a subsystem that stores or processes some resource-like quantity, while each directed link carries a positive weight specifying capacity, flow rate, or cost. The following formulation, adapted from the buffer-based model of \cite{Rantzer2018} yet reframed to underscore resource transfers, offers a unified lens for studying stability and control in large interconnected systems.

\subsection{Mathmatical Model}
Power grids, manufacturing lines, and logistics systems can all be viewed through the prism of resource flow networks, where each node accumulates or transforms a commodity and each directed edge imposes capacity or efficiency constraints \cite{Rantzer2018,Sala2019}. The broader category of \emph{complex networks} arises whenever the topology—potentially small-world, scale-free, or a hybrid—strongly affects operational outcomes \cite{Watts1998,Barabasi1999,Newman2010}. Node buffers, edge capacities, and nonnegativity requirements motivate the \emph{positive linear systems} framework \cite{Rantzer2018}, ensuring that stored quantities never become negative.

Small-world designs reduce average path lengths, potentially speeding up distribution but increasing the likelihood of regional bottlenecks. Scale-free graphs, dominated by a handful of hubs, exhibit robustness under random failures yet suffer if key hubs overload \cite{Albert2000}. Many real systems combine these traits \cite{Barthelemy2011}, creating hybrid effects on overall stability. Quantifying how different structural patterns translate to distinct global performance characteristics remains fundamental when developing control policies that preserve nonnegative resource flows.

Minor modifications in network connectivity, such as inserting shortcuts or reassigning link weights, can shift a system’s feasible region. A single overloaded node may trigger a domino of inefficiencies. By generalizing the buffer viewpoint of \cite{Rantzer2018} to emphasize structural versatility, one captures both high-level topological influences and the potential for significant interactions across the network. This perspective elucidates how each topology interacts with flow constraints, guiding more effective allocation and enhancing resilience in large-scale resource networks.

\subsection{Directed Graph and Weighted Adjacency}
Let 
\[
\widehat{\mathcal{G}} \;=\; (\mathcal{V}, \mathcal{E}, \mathcal{S}),
\]
where $\mathcal{V} = \{ v_1, \dots, v_n\}$ is the node collection, $\mathcal{E} \subseteq \mathcal{V}\times\mathcal{V}$ is the directed edge set, and each link $e_\ell$ has a positive weight $s_{e_\ell}$. Writing $s_{ij}$ for the weight of $(i,j)$, the adjacency matrix $M_{\widehat{\mathcal{G}}}\in\mathbb{R}^{n\times n}$ is given by
\[
[M_{\widetilde{\mathcal{G}}}]_{ij}
\,=\,
\begin{cases}
w_{ji}, &\text{if}\;(j,i)\in\mathcal{E},\\
0,      &\text{otherwise}.
\end{cases}
\]
For node $i$, its in-neighborhood is $\mathcal{N}_i^\mathrm{in}=\{\,j\mid (j,i)\in\mathcal{E}\}$, and its out-neighborhood is $\mathcal{N}_i^\mathrm{out}=\{\,j\mid (i,j)\in\mathcal{E}\}$.  

Following \cite{Rantzer2018}, suppose there is an origin subset $\mathcal{V}_o=\{1,\dots,|\mathcal{V}_o|\}$ with no incoming edges and a destination subset $\mathcal{V}_d$ with no outgoing edges, creating an overall flow from external sources to end sinks.

\subsection{Dynamics of Resource States}
Let $y_i(t)$ denote the resource content stored in node $i$ at time $t$. Define $g_i^\mathrm{in}$ and $g_i^\mathrm{out}$ as external input and output flows, respectively, while $v_{ij}$ models the transfer rate from node $i$ to $j$. Inspired by \cite{Rantzer2018}, the flow balances are
\[
\Sigma: \frac{dy_i}{dt}
=
\begin{cases}
g_i^\mathrm{in} - \sum_{j\in\mathcal{N}_i^\mathrm{out}} v_{ij}, & i\in\mathcal{V}_o,\\[4pt]
\displaystyle \sum_{j\in\mathcal{N}_i^\mathrm{in}}v_{ji}- \sum_{j\in\mathcal{N}_i^\mathrm{out}}v_{ij}, & i\notin\mathcal{V}_o\cup\mathcal{V}_d,\\[4pt]
\displaystyle \sum_{j\in\mathcal{N}_i^\mathrm{in}}v_{ji} - g_i^\mathrm{out}, & i\in\mathcal{V}_d.
\end{cases}
\]
Hence, an origin node acquires external input and forwards resources to out-neighboring nodes, while a destination node collects inflows but expends $g_i^\mathrm{out}$. Intermediate nodes simply redistribute resources in both directions.

For stability analysis, we assume linear flows of the form
\[
g_i^\mathrm{out}=\kappa_i\,y_i, 
\qquad
v_{ij}=\alpha_{ij}\,s_{ij}\,y_i,
\]
where $\{\kappa_i\}_{i\in\mathcal{V}_d}$ and $\{\alpha_{ij}\}_{(i,j)\in\mathcal{E}} \in \alpha$ are tuning parameters. Each edge $(i,j)$ thus has a distinct $\alpha_{ij}$, allowing a nonuniform distribution of flows throughout the network. This setup generalizes simpler node-based models by granting edges explicit freedom to modulate their individual flow behavior, ensuring a broad range of potential configurations.
\noindent
Let \(L_1\colon \mathbb{R}_+ \to \mathbb{R}_+^n\) denote the space of Lebesgue-integrable functions on \(\mathbb{R}_+\). For any \(f \in L_1\), define
\[
\|f\|_{L_1} 
\;=\; 
\int_{0}^{\infty} \|f(t)\|_1 \,dt.
\]
Suppose that the system \(\Sigma\) is internally stable, and there exists some \(\gamma > 0\) such that \(\|x\|_{L_1} \le \gamma \,\|\nu\|_{L_1}\) for every \(\nu \in L_1\). In that case, the \(L_1\)-gain of \(\Sigma\), denoted by \(\|\Sigma\|_1\), is given by
\[
\|\Sigma\|_1
\;=\;
\sup_{\nu \in L_1}
\frac{\|x\|_{L_1}}{\|\nu\|_{L_1}}.
\]

\begin{prob}\label{pb:1}
Let \(\bar{L}\) be the target cost for tuning the parameter \(\alpha_{ij}\). The objective is to choose \(\alpha_{ij} \in \Theta\) so as to minimize the \(L_1\)-gain of \(\Sigma\), subject to a cost constraint on \(\alpha_{ij}\). Concretely,
\begin{equation}
\begin{aligned}
\min_{\alpha_{ij} \in \Theta}
\quad
&\|\Sigma_\theta\|_1
\\
\text{subject to}
\quad
&\Psi(\alpha)\;\le\;\bar{\Psi},\\
&\Sigma \;\text{is internally stable},
\end{aligned}
\end{equation}
where \(\alpha_{ij}\) is the parameter to be determined, and \(\Psi\colon \Theta \to [0,\infty)\) is a nonnegative cost function representing the resources required to implement \(\alpha_{ij}\).
\end{prob}

\section{Centrality Measurements for Topology Networks}
\subsection{Katz Centrality}

Katz centrality is a classical centrality measure used to evaluate the influence of nodes in a network by considering both the direct and indirect connections. Given a directed graph \( G = (V, E) \) with adjacency matrix \( A \in \mathbb{R}^{n \times n} \), the Katz centrality vector \( \mathbf{r} \in \mathbb{R}^{n} \) is defined as the unique solution to the following linear equation:
\begin{equation}
    \mathbf{r} = \alpha A \mathbf{r} + \mathbf{1},
\end{equation}
where \( \alpha \in (0, \frac{1}{\rho(A)}) \) is a positive attenuation factor satisfying the convergence condition, \( \rho(A) \) denotes the spectral radius of \( A \), and \( \mathbf{1} \in \mathbb{R}^{n} \) is a vector of ones.

Equivalently, Katz centrality can be written in the closed-form expression:
\begin{equation}
    \mathbf{r} = (I - \alpha A)^{-1} \mathbf{1}.
\end{equation}
This expression shows that Katz centrality accounts for all walks in the network, where longer walks are exponentially downweighted by powers of \( \alpha \).

In terms of control-theoretic interpretation, Katz centrality can also be understood as the steady-state solution of a continuous-time linear dynamical system:
\begin{equation}
    \frac{d\mathbf{x}}{dt} = \alpha A \mathbf{x} + \mathbf{1} - \mathbf{x},
\end{equation}
where the term \( \mathbf{1} \) denotes uniform external input, \( \alpha A \mathbf{x} \) models the influence from neighboring nodes, and \( -\mathbf{x} \) ensures stability. At steady-state, the dynamics yield the same solution as in Eq. (2).

In practical computations, especially for large-scale or dynamic graphs, iterative methods such as Jacobi or power iteration are employed to approximate the solution to Eq. (1) due to the high computational cost of matrix inversion. This makes Katz centrality particularly suitable for scalable applications in network science and optimization.

\subsection{Eigenvector Centrality}

Eigenvector centrality is a spectral measure of node importance in a network, which accounts not only for the number of direct connections a node has, but also for the quality of those connections. Let $G = (V, E)$ be an undirected network with $n$ nodes, and let $A \in \mathbb{R}^{n \times n}$ denote its adjacency matrix. The eigenvector centrality vector $x \in \mathbb{R}^n$ is defined as the solution of the following eigenvalue problem:
\begin{equation}
    A x = \lambda x,
\end{equation}
where $\lambda$ is the largest eigenvalue of $A$. The $i$-th entry $x_i$ of the vector $x$ represents the centrality of node $i$, and it satisfies:
\begin{equation}
    x_i = \frac{1}{\lambda} \sum_{j=1}^{n} A_{ij} x_j.
\end{equation}
This implies that a node is central if it is connected to other nodes that are themselves central. Unlike degree centrality, which treats all neighbors equally, eigenvector centrality assigns higher scores to nodes that are connected to other high-scoring nodes.

This measure is particularly suitable in applications where influence or importance propagates through the network recursively. It is widely used in social network analysis, biological systems, and control of complex networks, and serves as the basis for PageRank and other link-based ranking algorithms.

\subsection{Closeness Centrality}

Let $G = (V, E)$ be a connected, undirected graph, where $V$ denotes the set of nodes with cardinality $n = |V|$, and $E$ is the set of edges. The shortest path distance between node $v \in V$ and another node $w \in V$ is denoted by $d(v, w)$.

Then, the closeness centrality of node $v$ is defined as:
\begin{equation}
C_{\mathrm{clo}}(v) = \frac{n-1}{\sum\limits_{w \in V, w \neq v} d(v, w)}
\label{eq:closeness}
\end{equation}

This metric quantifies how efficiently information can be propagated from node $v$ to all other nodes in the network. A higher value of $C_{\mathrm{clo}}(v)$ indicates that node $v$ is, on average, closer to other nodes and can thus reach them more quickly.

\subsection{Betweenness Centrality}

Betweenness centrality is a widely used index for quantifying the influence of a node in a network in terms of the control it exerts over information flow. Given a graph $\mathcal{G} = (\mathcal{V}, \mathcal{E})$ with $|\mathcal{V}| = n$ nodes, the betweenness centrality $C_B(v)$ of a node $v \in \mathcal{V}$ is defined as follows:
\begin{equation}
C_B(v) = \sum_{\substack{s,t \in \mathcal{V}\\ s \ne t \ne v}} \frac{\sigma_{st}(v)}{\sigma_{st}},
\label{eq:betweenness}
\end{equation}
where $\sigma_{st}$ denotes the total number of shortest paths from node $s$ to node $t$, and $\sigma_{st}(v)$ denotes the number of those paths that pass through node $v$.

The betweenness centrality can be normalized by dividing it by the maximum possible value:
\begin{equation}
C_B^{\text{norm}}(v) = \frac{2}{(n-1)(n-2)} C_B(v), \quad \text{for } n > 2.
\end{equation}

Betweenness centrality reflects the importance of a node as an intermediary, and is particularly relevant in assessing node vulnerability or control over communication in distributed and social systems.

\subsection{PageRank Centrality}
PageRank centrality evaluates the influence of nodes in a directed network by computing the stationary distribution of a random walk with restarts. Consider a network with $n$ nodes and let ${P} \in \mathbb{R}^{n \times n}$ be a column-stochastic matrix representing the transition probabilities between nodes. The PageRank vector ${x} \in \mathbb{R}^n$ is defined as the unique solution to the linear system
\begin{equation}
({I} - \alpha {P}) {x} = (1 - \alpha){v},
\end{equation}
where $\alpha \in (0,1)$ is a damping parameter and ${v}$ is a personalization vector with ${1}^\top {v} = 1$.

This formulation models a stochastic process in which, at each step, the walker follows outgoing links with probability $\alpha$ or teleports to a random node with probability $1 - \alpha$. The resulting vector ${x}$ assigns higher scores to nodes that are frequently visited in this process, thereby capturing both local connectivity and global reachability within the network.

\section{Simulation Experiments and Analysis}

\subsection{Practical Guidelines for Metric Selection}
\label{subsec:practical}
Table~\ref{tab:guideline} summarises heuristic links between
infrastructure traits and centrality choice.
\begin{table}[h]
  \centering
  \caption{Heuristic metric selection for typical networks}
  \label{tab:guideline}
  \begin{tabular}{@{}lll@{}}
 \toprule
    Scenario & Dominant risk & Suggested metric \\ \midrule
    HV grid (hub dominated) & Cascading via hubs & Betweenness / Eigenvector \\
    Urban transport (meshed) & Local congestion  & Closeness \\
    Gas pipeline (tree‑like) & Source isolation   & Katz       \\
    Internet backbone        & Traffic ranking    & PageRank   \\ \bottomrule
  \end{tabular}
\end{table}

These qualitative rules are later corroborated by quantitative
experiments, providing practitioners with an
\emph{a‑priori} rationale before expensive simulation runs.

\subsection{Experimental Setup}
\begin{figure}
    \centering
\includegraphics[width=1\linewidth]{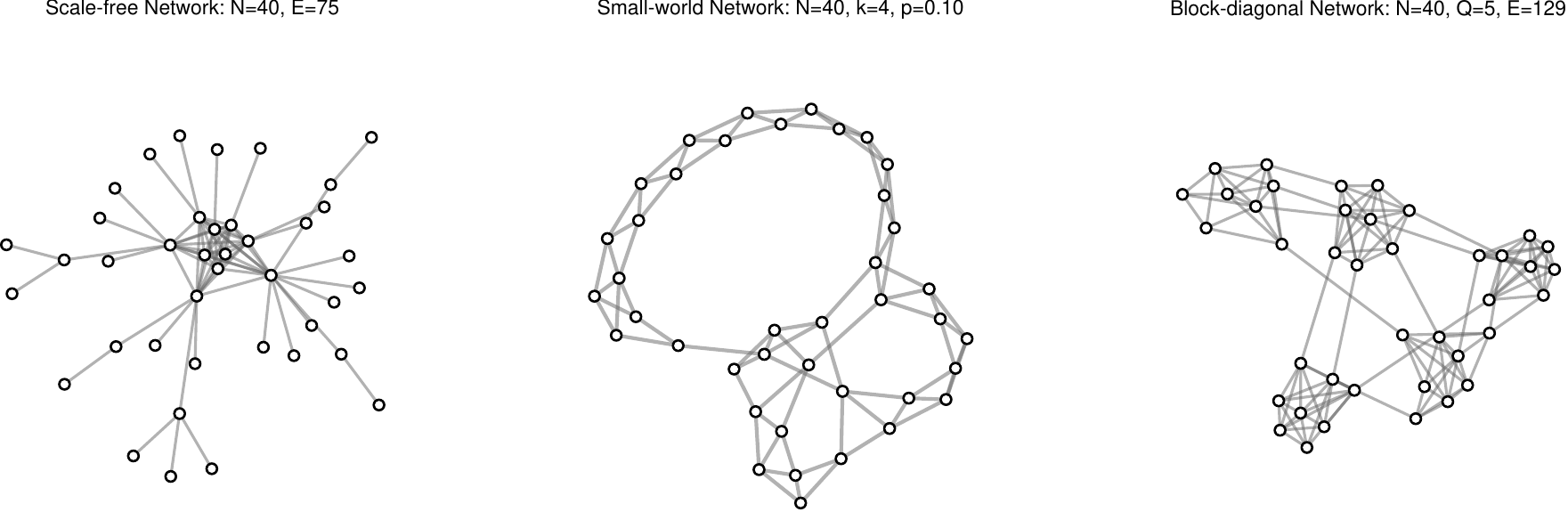}
    \caption{Randomly generated example networks of Scale-free network, Small-world network, and Block-diagonal network with $40$ nodes.}
    \label{fig:1}
\end{figure}

To systematically investigate how different node-centric strategies affect stability in energy-flow networks, we focus on a small-world topology that balances local clustering with relatively short path lengths. For each simulation, a small-world graph with $n$ nodes is generated by starting from a ring lattice and rewiring edges at a probability $p\in(0,1)$, producing variations in structural properties. Each node’s initial energy $x_i(0)$ is drawn uniformly from $[0,1]$, while node- and edge-level parameters $(\beta_i,\delta_{ij})$ are randomly assigned within $[0,0.5]$ to reflect diverse transfer capacities. Five primary centrality measures (betweenness, closeness, eigenvector, Katz, and PageRank) determine node “importance.” A set fraction of high-ranked nodes under each metric receives additional resources—e.g., gain settings or buffer capacity—aimed at minimizing the global $L_1$ norm of the energy trajectory. This procedure is repeated for $M$ randomly generated network instances, allowing us to compute ensemble statistics across various small-world realizations. Fig.~\ref{fig:1} shows the examples of randomly generated energy flow networks for further analysis.

\subsection{Results Analysis}

\begin{figure}
    \centering
\includegraphics[width=1\linewidth]{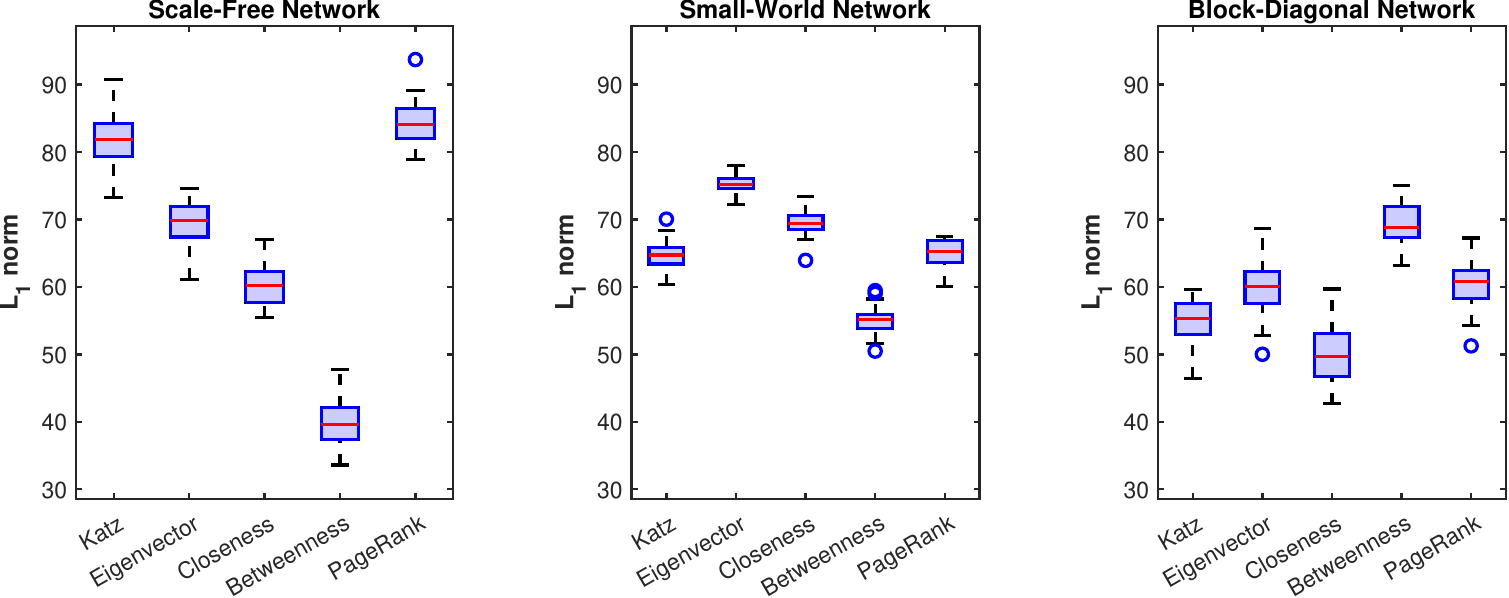}
    \caption{The optimized $L_1$-norm of energy flow network $\Sigma$ under Scale-free network, Small-world network, and Block-diagonal network with different centrality measurements.}
    \label{fig:2}
\end{figure}

Figure~\ref{fig:2} presents the box plots of the $L_1$-norm simulations for each centrality-based allocation strategy across three distinct network topologies—Scale-Free, Small-World, and Block-Diagonal. In our experiments, each box plot summarizes the median, interquartile range, and outliers, thereby providing a comprehensive visual comparison of the performance and robustness of the different centrality measures under varying networks.
For the Scale-Free networks, the results indicate that betweenness and eigenvector centrality measures consistently achieve lower $L_1$-norm. This finding suggests that in networks characterized by hub-dominated structures, nodes that either act as bridges (high betweenness) or possess significant global influence (high eigenvector values) are more effective at optimizing resource distribution or dampening disruptive flows. Conversely, closeness-based allocations tend to exhibit higher $L_1$-norm, which could be attributed to the fact that relying solely on shortest path distances may neglect the influence of critical hub nodes that are pivotal in maintaining the network’s overall stability.

In the case of Small-World networks, the distribution of $L_1$-norm becomes somewhat narrower, reflecting the more homogeneous connectivity inherent in such networks. Here, while betweenness and eigenvector measures still outperform others, the differences between centrality metrics are less pronounced compared to the Scale-Free case. This observation underscores that the overall network structure—characterized by high clustering and short path lengths—tends to equalize the impact of different allocation strategies to some extent.

For Block-Diagonal networks, which inherently possess community structures with relatively sparse inter-module connections, the box plots reveal a distinct behavior. In these networks, betweenness centrality again performs well, highlighting the importance of inter-community bridges. However, the variability in the $L_1$-norm is higher compared to the Small-World networks, likely due to the uneven distribution of connections between modules. Katz and PageRank measures offer competitive performance in all three network types, although their effectiveness appears to be more sensitive to changes in the rewiring probability 
$p$ increases, alterations in local connectivity patterns seem to influence the performance of these measures, leading to variations in the corresponding $L_1$-norm.

Overall, the experimental results affirm that the choice of centrality metric has a significant impact on both the median performance and the robustness of the allocation strategy. In energy-flow networks or similar systems, selecting a centrality measure that effectively captures the roles of bridging nodes or high-impact hubs can markedly enhance stability and performance. These insights align with our initial hypotheses and underscore the importance of considering network topology when designing centrality-based allocation strategies.

\subsection{Monte‑Carlo Protocol}
All simulations are averaged over $M=200$ random graphs.
Box‑plots in Fig.~\ref{fig:2} show medians and
\emph{95\%} bootstrap confidence intervals (10\,000 resamples).
Performance differences are evaluated with a two‑sided Wilcoxon
signed‑rank test; an asterisk ($^\ast$) denotes $p<0.05$.

\section{conclsion}
In this paper, we have investigated how different centrality-based strategies affect the stability optimization of energy-flow networks. By constructing a generalized dynamical model and applying five widely used centrality measures—Katz, eigenvector, closeness, betweenness, and PageRank—we evaluated their impact on minimizing the $L_1$-norm across various network topologies, including scale-free, small-world, and block-diagonal structures. Our simulation results demonstrated that the effectiveness of each centrality measure strongly depends on the underlying network structure. Notably, betweenness and eigenvector centralities showed superior performance in hub-dominated networks, while differences were less pronounced in more homogeneous structures such as small-world networks. These findings emphasize the importance of selecting appropriate centrality metrics tailored to the network's structural properties and provide practical guidance for designing robust and efficient control strategies in large-scale resource flow systems.


\begin{thebibliography}{99}

\bibitem{1}
Duan, Z., Wang, J., Chen, G., Huang, L.: Stability analysis and decentralized control of a class of complex dynamical network. Automatica \textbf{44}(4), 1028--1035 (2008)

\bibitem{2}
Xiang, J., Chen, G.: On the V-stability of complex dynamical networks. Automatica \textbf{43}(6), 1049--1057 (2007)

\bibitem{34}
Mei, W., Efimo, D., Ushirobira. R.: On input-to-output stability and robust synchronization of generalized Persidskii systems. IEEE Trans. Autom. Control \textbf{67}(10), 5578--5585 (2021)


\bibitem{3}
Wang, L., Dai, G.: Global stability of virus spreading in complex heterogeneous networks. SIAM J. Appl. Math. \textbf{68}(5), 1495--1502 (2008)


\bibitem{33}
Ogura, M., Mei, W., Sugimoto, K.: Synergistic effects in networked epidemic spreading dynamics. IEEE Trans. Circuits Syst. II Express Briefs \textbf{67}(3), 496--500 (2019)


\bibitem{4}
Nowzari, C., Preciado, V.M., Pappas, G.J.: Analysis and control of epidemics: A survey of spreading processes on complex networks. IEEE Control Syst. Mag. \textbf{36}(1), 26--46 (2016)

\bibitem{5}
Chu, C.C., Iu, H.H.C.: Complex networks theory for modern smart grid applications: A survey. IEEE J. Emerg. Sel. Top. Circuits Syst. \textbf{7}(2), 177--191 (2017)

\bibitem{6}
Caccioli, F., Catanach, T.A., Farmer, J.D.: Heterogeneity, correlations and financial contagion. Adv. Complex Syst. \textbf{15}(2), 1250058 (2012)

\bibitem{7}
Battiston, S., Caldarelli, G., May, R.M., Roukny, T., Stiglitz, J.E.: The price of complexity in financial networks. Proc. Natl. Acad. Sci. USA \textbf{113}(36), 10031--10036 (2016)

\bibitem{8}
Li, Z., Chen, G.: Global synchronization and asymptotic stability of complex dynamical networks. IEEE Trans. Circuits Syst. II Express Briefs \textbf{53}(1), 28--33 (2006)

\bibitem{9}
Vu, T.L., Turitsyn, K.: A framework for robust assessment of power grid stability and resiliency. IEEE Trans. Autom. Control \textbf{62}(3), 1165--1177 (2016)

\bibitem{10}
Krishnan, V., Kastrouni, E., Pyrialakou, V.D., Gkritza, K., McCalley, J.D.: An optimization model of energy and transportation systems: Assessing the high-speed rail impacts in the United States. Transp. Res. Part C Emerg. Technol. \textbf{54}, 131--156 (2015)

\bibitem{11}
Barucca, P., Bardoscia, M., Caccioli, F., D'Errico, M., Visentin, G., Caldarelli, G., Battiston, S.: Network valuation in financial systems. Math. Finance \textbf{30}(4), 1181--1204 (2020)

\bibitem{12}
Liu, Y.-Y., Slotine, J.-J., Barabási, A.-L.: Controllability of complex networks. Nature \textbf{473}(7346), 167--173 (2011)

\bibitem{13}
Ghoshal, G., Barabási, A.L.: Ranking stability and super-stable nodes in complex networks. Nat. Commun. \textbf{2}(1), 394 (2011)

\bibitem{14}
Ruths, J., Ruths, D.: Control profiles of complex networks. Science \textbf{343}(6177), 1373--1376 (2014)

\bibitem{15}
Katz, L.: A new status index derived from sociometric analysis. Psychometrika \textbf{18}(1), 39--43 (1953)


\bibitem{16}
Bonacich, P.: Some unique properties of eigenvector centrality. Soc. Netw. \textbf{29}(4), 555--564 (2007)

\bibitem{17}
Okamoto, K., Chen, W., Li, X.Y.: Ranking of closeness centrality for large-scale social networks. Int. Workshop on Frontiers in Algorithmics, pp. 186--195. Springer, Berlin, Heidelberg (2008)

\bibitem{18}
Brandes, U.: A faster algorithm for betweenness centrality. J. Math. Sociol. \textbf{25}(2), 163--177 (2001)

\bibitem{19}
Xing, W., Ghorbani, A.: Weighted pagerank algorithm. In: Proc. 2nd Annual Conf. on Communication Networks and Services Research, pp. 305--314. IEEE (2004)

\bibitem{20}
Beard, C.C., Frost, V.S.: Prioritized resource allocation for stressed networks. IEEE/ACM Trans. Netw. \textbf{9}(5), 618--633 (2001)


\bibitem{21}
Wang, W.X., Ni, X., Lai, Y.C., Grebogi, C.: Optimizing controllability of complex networks by minimum structural perturbations. Phys. Rev. E \textbf{85}(2), 026115 (2012)





\bibitem{22}
Sun, Y.Z., Leng, S.Y., Lai, Y.C., Grebogi, C., Lin, W.: Closed-loop control of complex networks: A trade-off between time and energy. Phys. Rev. Lett. \textbf{119}(9), 198301 (2017)

\bibitem{23}
Borgatti, S.P.: Centrality and network flow. Social Networks \textbf{27}(1), 55--71 (2005)

\bibitem{24}
Prell, C., Hubacek, K., Reed, M.: Stakeholder analysis and social network analysis in natural resource management. Society and Natural Resources \textbf{22}(6), 501--518 (2009)




\bibitem{Mei2025}
Mei, W., Zheng, D., Zhou, Y., Taha, A., Zhao, C.: On input-to-state stability verification of identified models obtained by Koopman operator. J. Franklin Inst. \textbf{362}(2), 107490 (2025)

\bibitem{Rantzer2018}
Rantzer, A., Valcher, M.E.: A tutorial on positive systems and large scale control. In: Proc. 57th IEEE Conf. Decision and Control, pp. 3686--3697 (2018)

\bibitem{Sala2019}
Beck, J., Kempener, R., Cohen, B., Petrie, J.: A complex systems approach to planning, optimization and decision making for energy networks. Energy Policy \textbf{36}(8), 2795--2805 (2008)

\bibitem{Watts1998}
Watts, D.J., Strogatz, S.H.: Collective dynamics of ‘small-world’ networks. Nature \textbf{393}(6684), 440--442 (1998)

\bibitem{Barabasi1999}
Barabási, A.-L., Albert, R.: Emergence of scaling in random networks. Science \textbf{286}(5439), 509--512 (1999)

\bibitem{Newman2010}
Newman, M.E.J.: Networks: An Introduction. Oxford University Press, Oxford (2010)

\bibitem{Albert2000}
Albert, R., Jeong, H., Barabási, A.-L.: Error and attack tolerance of complex networks. Nature \textbf{406}(6794), 378--382 (2000)

\bibitem{Barthelemy2011}
Barthélemy, M.: Spatial networks. Phys. Rep. \textbf{499}(1--3), 1--101 (2011)



\end{thebibliography}
\end{document}